\documentclass[epj]{svjour}
\usepackage[utf8]{inputenc}
\usepackage{graphicx,xcolor}
\usepackage{amsmath,amssymb}
\usepackage{multirow}

\title{A simplified estimate of the Effective Reproduction Number $R_t$ using its relation with the doubling time and application to Italian COVID-19 data}
\author{
Gianluca Bonifazi\inst{1,2} \and 
Luca Lista\thanks{Corresponding author, e-mail: {\tt luca.lista@infn.it}.} \inst{, 3, 4} \and
Dario Menasce\inst{5} \and
Mauro Mezzetto\inst{6} \and
Daniele Pedrini\inst{5} \and
Roberto Spighi\inst{2} \and
Antonio Zoccoli\inst{7, 2}
}
\institute{
Universit\`a Politecnica delle Marche \and 
INFN Sezione di Bologna \and
Universit\`a degli Studi di Napoli Federico II \and
INFN Sezione di Napoli \and
INFN Sezione di Milano Bicocca \and
INFN Sezione di Padova \and
Alma Mater Studiorum Universit\`a di Bologna
}

\date{Received: date / Revised version: date}

\abstract{
A simplified method to compute $R_t$, the Effective Reproduction Number, is presented. The method relates the value of $R_t$ to the estimation of the doubling time performed with a local exponential fit. The condition $R_t=1$ corresponds to a growth rate equal to zero or equivalently an infinite doubling time. Different assumptions on the probability distribution of the generation time are considered. A simple analytical solution is presented in case the generation time follows a gamma distribution.
}

\begin{document}
\maketitle
\section{Introduction}
The Effective Reproduction Number $R_t$ is one of the main parameters that controls the evolution of an infection. It recently gained importance during the COVID-19 pandemic outbreak and is used as one of the indicators to determine restrictive measures such as regional or national lock-downs.

Different algorithms for its computation are available~\cite{wallinga}~\cite{bettencourt}~\cite{cori}~\cite{rki}~\cite{systrom}, some of which are very CPU intensive. 

Implementations are also available as software packages~\cite{epiestim} for a number of algorithms, and results are presented on websites~\cite{rt-live}~\cite{rt-italy}~\cite{our web site} with regular updates.

CPU-effective algorithms offer the advantage that estimates can be derived in real time as soon as new data are published. Often, results of simplified algorithms don't differ too much from the results of more accurate methods, in particular due to the limited quality of input data.

The following proposes a simplified approach to the estimate of $R_t$ based on a determination of the doubling time, or equivalently the growth rate, which can be simply achieved with a regression procedure.

\section{The Effective Reproduction Number, $R_t$}
We assume $I_t$ is the number of infected persons at the time
$t$, measured as number of days from a conventional beginning of the epidemic, defined as $t=0$.

Each contagious person can infect other people during his infection period. We assume that a person that got infected at a day $d$ will infect, on average, a certain number of other persons that become infectious at the day $t>d$ with a discrete probability distribution $w_{s}$, with $s=t-d$. The newly infected people, on turn, may infect more people with the same mechanism.
$s=t-d$ is defined as the generation time in literature and corresponds to the time interval between infector-infected pair.

The probability distribution $w_{s}$ is normalized to unity:
\begin{equation}
    \sum_{s=1}^{\infty}w_s = 1\,.
\end{equation}
In practice, after a sufficiently large amount of time, {\it i.e.}: for a sufficiently large value of $s$, $w_{s}$ becomes negligible.
An estimate of $w_s$ from Italian infection data, unfortunately from a limited number of cases, is published in~\cite{cereda} where $w_s$ is approximated with a gamma distribution.

At a time $t$, the expected number of infected persons, $\mathbb{E}[I_{t}]$ can be determined from $I_{d}$, $d=0, \cdots, t-1$, according to~\cite{cori}, as:
\begin{equation}
\mathbb{E}[I_{t}] = R_t \sum_{d=0}^{t-1} I_{d}w_{t-d}\,,
\end{equation}
or, equivalently, defining $s=t-d$, as:
\begin{equation}
\mathbb{E}[I_{t}] = R_t \sum_{s=1}^{t} I_{t-s}w_{s}\,.
\label{eq:eit}
\end{equation}

The simplest assumption on $w_s$ is a constant generation time $g$, which is equivalent $w_s = \delta_{gs}$ where $\delta_{gs}$ is a Kronecker delta, {\it i.e.}:  $w_g=1$ and $w_s=0$ for $s\ne g$.
In this case, Eq~\ref{eq:eit} becomes:
\begin{equation}
\mathbb{E}[I_{t}] = R_t I_{t-g}\,.
\label{eq:const_t_growth}
\end{equation}
%
%
For COVID-19, the average generation time, defined as the mean value of a gamma distribution fitted to the Italian data, is $g=6.7 \pm 1.9$ days ~\cite{cereda}. The Robert Koch Institute (RKI) takes instead for Germany the value $g=4$ that gives a very simple estimate $\hat{R}_t$ of $R_t$~\cite{rki} \footnote{We tested this algorithm over the $R_t$ values computed by RKI for the German cases as reported in \cite{Excel RKI}, and we found an excellent agreement with the data published by the RKI.}:
\begin{equation}
    \hat{R}_t = \frac{I_t}{I_{t-g}}\,,
\end{equation}
or the smoother ratio of the moving averages over $g$ days:
\begin{equation}
    \hat{R}_t = \frac{\sum_{d=t-g+1}^t I_d}{\sum_{d=t-g+1}^t I_{d-g}}\,.
\label{eq:rt_rki}
\end{equation}

Usually, the moving average over few days does not sufficiently smooth  the distribution of the number of daily infected cases $I_t$. In particular, the lower number of swab tests taken during the weekend causes a ``ripple'' structure that requires a further smoothing to be applied to the input data before evaluating Eq.~\ref{eq:rt_rki}.

\begin{figure}[htbp]
    \centering
    \includegraphics[width=0.9\textwidth]{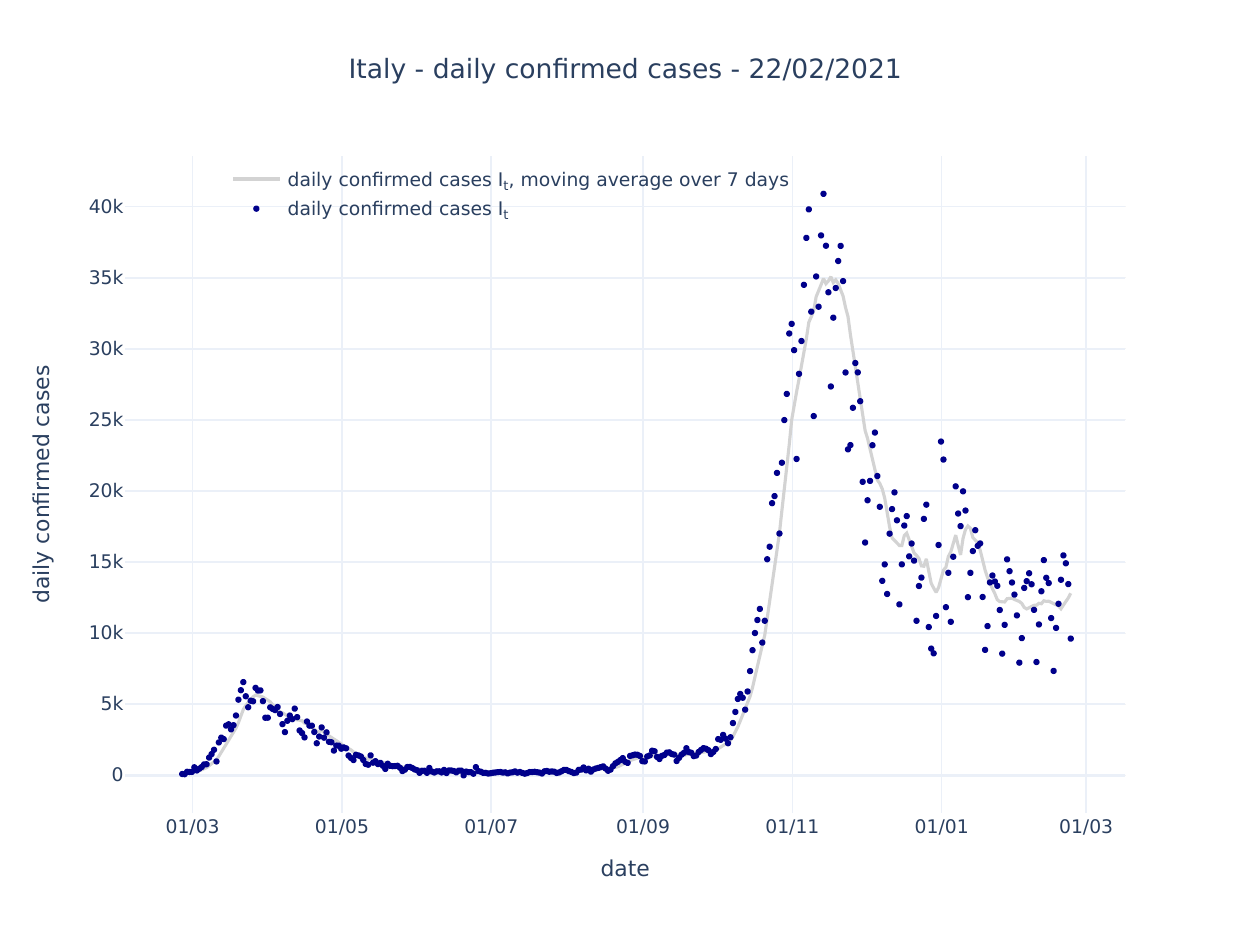}
    \caption {Number of daily confirmed cases, $I_t$ for Italy according to public COVID-19 Italian data from the Italian Dipartimento di Protezione Civile, dark blue dots. The moving average over 7 days is also shown as light gray line.}
    \label{fig:italy_pos}
\end{figure}

Figure~\ref{fig:italy_pos} shows the number of daily confirmed cases, $I_t$ for Italy according to public COVID-19 Italian data from the Italian Dipartimento di Protezione Civile. The large dispersion of data is clearly visible, in particular around the more stable moving average over 7 days, also shown in the figure.

\section{Relation between $R_t$ and doubling time}

Another indicator of the growth of the epidemic is the doubling time $\tau_2$ defined as the time required to double the number of infected persons, assuming an exponential growth.

Given $n$ consecutive counts of infected people, $I_{t-n+1}, \cdots, I_t$, the following function model can interpolate the $n$ counts:
\begin{equation}
    I_t = A\,e^{\lambda t}\,,
    \label{eq:exp_fit_model}
\end{equation}
or, equivalently:
\begin{equation}
    I_t = A\,2^{t/\tau_2}\,.
    \label{eq:exp_fit_model_2}
\end{equation}
The growth rate $\lambda$ is related to the doubling time $\tau_2$ by:
\begin{equation}
\tau_2 = \frac{\log{2}}{\lambda}\,.
\label{eq:doubling_time}
\end{equation}

Estimates of $A$ and $\lambda$, or equivalently $\tau_2$, can be determined with a numerical fit procedure. In particular, the exponential fit can be conveniently implemented as a linear regression on $\log{I_t}$.

Assuming $R_t=R$ constant during the considered time interval, the evolution model in Eq.~\ref{eq:const_t_growth} represents an exponential growth. In a time period formed by a number of days $n$ which is an integer multiple of $g$: $n=N g$, we have:
\begin{equation}
\mathbb{E}[I_{t}] = R^{\,N} I_{h}\,,
\end{equation}
where $h=t-n$, or:
\begin{equation}
\mathbb{E}[I_{t}] = R^{\,n/g} I_{h}\,.
\end{equation}
Changing the base from $R$ to $e$ gives:
\begin{equation}
\mathbb{E}[I_{t}] = e^{(n\log{R})/g} I_{h}\,.
\end{equation}
Comparing with Eq.~\ref{eq:exp_fit_model}, considering that $t=h+n$, and $A\,e^{\lambda t} = A\,e^{\lambda n}e^{\lambda h}$,  we have:
\begin{equation}
    \lambda = \frac{\log R}{g}\,,\,\,\,\ A={I_{h}}e^{-\lambda h}\,,
\end{equation}
hence the estimate $\hat{R}$ of $R$ is:
\begin{equation}
    \hat{R} = e^{g\hat{\lambda}} = e^{(g\log{2})/\hat{\tau}_2}\,,
    \label{eq:r_e_g_lambda}
\end{equation}
where $\hat{\lambda}$ and $\hat{\tau}_2$ are the estimates of $\lambda$ and ${\tau}_2$, respectively.

\section{Simplified algorithm}
We studied the progression of the COVID-19 pandemic in Italy, considering the data published on daily basis by Italian Dipartimento di Protezione Civile \cite{dpc}.
For each day $t$, we perform an exponential fit to the $n$ last days' counts, $I_{t-n+1}$, $ \cdots,$ $I_t$. We determine an estimate $\hat{\tau}_2$ of the doubling time $\tau_2$, or an equivalent estimate  $\hat{\lambda}$ of the growth rate $\lambda$, from a fit of the model in Eq.~\ref{eq:exp_fit_model} or Eq.~\ref{eq:exp_fit_model_2}. Then, assuming a reasonable estimate $g$ of the average generation time, we estimate $R_t$ according to Eq.~\ref{eq:r_e_g_lambda} as:
\begin{equation}
    \hat{R}_t = e^{g\hat{\lambda}} = e^{(g\log{2})/{\hat{\tau}_2}} = 2^{g/\hat{\tau}_2}\,.
    \label{eq:rt_lista}
\end{equation}

There are some advantages of Eq.~\ref{eq:rt_lista} compared to the simplified model from Eq.~\ref{eq:rt_rki}:
\begin{itemize}
    \item Eq.~\ref{eq:rt_lista} can also be applied in case $g$, the average generation time, is not an integer, while Eq.~\ref{eq:rt_rki} must approximate $g$ to the nearest integer.
    \item The exponential fit better follows an exponential growth in the considered time interval, as it is the case when $R_t$ is a constant, with respect to a moving average.
\end{itemize}
At the cost of a modest increase in the computing time, yet maintaining very good speed, we consider the method proposed here to be more flexible and reliable compared to the method adopted in~\cite{rki}. Moreover, the data smoothing can be tuned by including a sufficient number of points in the fit. In this way, no preliminary smoothing of the data is needed before the application of the algorithm.

In the following sections, we will introduce extensions of Eq.~\ref{eq:rt_lista} that allow a more precise determination of $\hat{R}_t$ than with the simplified assumption that $w_s=\delta_{gs}$, {\it i.e.}: $s$ is constant and equal to $g$.

\section{Uncertainty estimate}
Given Eq.~\ref{eq:rt_lista}, the uncertainty on $\hat{R}_t$ is determined by the uncertainties on $\hat{\lambda}$ (or $\tau_2$) and the uncertainty on $g$. Namely, if $\sigma_{\hat{\lambda}}$ and $\sigma_g$ are the uncertainties on $\hat{\lambda}$ and $g$, respectively, within a Gaussian error approximation, the variance of $R_t$ is given by:
\begin{equation}
    \mathrm{Var}[\hat{R}_t] = \left(\frac{\partial \hat{R}_t}{\partial \hat{\lambda}}\sigma_{\hat{\lambda}}\right)^2+\left(\frac{\partial \hat{R}_t}{\partial g}\sigma_g\right)^2 =
    (g e^{\hat{\lambda} g}\sigma_{\hat{\lambda}})^2+(\hat{\lambda} e^{\hat{\lambda} g}\sigma_g) =
    (e^{\hat{\lambda} g})^2 (g^2\sigma_{\hat{\lambda}}^2 + \hat{\lambda}^2\sigma_g^2)\,.
\end{equation}
The error on $\hat{R}_t$ is:
\begin{equation}
    \sigma_{\hat{R}_t} = \sqrt{\mathrm{Var}[\hat{R}_t] } =
    \hat{R}_t \sqrt{g^2\sigma_{\hat{\lambda}}^2 + \hat{\lambda}^2\sigma_g^2}\,.
\label{eq:rt_err}
\end{equation}

The uncertainty on $\hat{\lambda}$ derives from the exponential fit procedure, while the uncertainty on $g$ depends on how well the probability distribution of the generation time $w_s$ is known. From~\cite{cereda}, the estimate of $w_s$ and its average $g$ for COVID-19 in Italy is known from a limited number of cases.

In particular, when $\hat{\lambda}=0$ (infinite doubling time), which corresponds to $\hat{R}_t=1$, $\sigma_g$ doesn't contribute to the $\hat{R}_t$ uncertainty. This means that an imperfect assumption on $g$ does not affect the condition $\hat{R}_t=1$ which is important to determine the turning point of infection, from growing to receding, or vice versa.

The uncertainty computed in Eq.~\ref{eq:rt_err} does not take into account the systematic uncertainty due to the assumed approximation that the generation time $s$ is constant, and equal to $g$.
Moreover, the assumption of Gaussian uncertainties may not hold for an asymmetric distribution.

\section{Effect of finite width in the $w_s$ distribution}

The deviation of $w_s$ from the hypothesis of a constant generation time $s=g$ may be approximately estimated in the continuum approximation. Eq.~\ref{eq:eit} for a continuous time variable $t$ may be rewritten as:
\begin{equation}
  i(t) = \rho(t)\int_0^t i(t-s)\,w(s)\,\mathrm{d}s\,,
  \label{eq:it_continuous}
\end{equation}
where $\rho(t)$ and $i(t)$ are the continuum equivalent of $R$ and $I_t$, respectively.

The normalization condition is:
\begin{equation}
    \int_0^{\infty}w(s)\,\mathrm{d}s = 1\,.
\end{equation}

If $s$ is a constant equal to $g$, we have $w(s)=\delta(s-g)$, where $\delta$ is a Dirac's delta function. Hence:
\begin{equation}
    i(t) = \rho(t)\,i(t-g)\,.
\end{equation}
Assuming an exponential growth $i(t) = A\,e^{\lambda t}$, one has:
\begin{equation}
    A\,e^{\lambda t} = \rho(t)\,A\,e^{\lambda(t-g)} = \rho(t)\,A\, e^{\lambda t}e^{-\lambda g}\,, 
\end{equation}
which gives the continuous version of Eq.~\ref{eq:rt_lista}, where $\rho(t)=\rho$ is a constant:
\begin{equation}
    \rho = e^{\lambda g}\,.
    \label{eq:rho_e_lambda_g}
\end{equation}

Assuming, instead, that $w(s)$ deviates from the Dirac's delta assumption and has average value $g$ and standard deviation $\sigma$, we may write Eq.~\ref{eq:it_continuous} applying a series expansion of $i(t-s)$ around $s=g$:
\begin{equation}
    i(t) = \rho(t)\int_0^t \left[ i(t-g)\,w(s) - i^\prime(t-g)\,(s-g)\,w(s)+ \frac{1}{2}i^{\prime\prime}(t-g)\,(s-g)^2\,w(s) +\cdots \right]\,\mathrm{d}s\,.
    \label{eq:rt_cont_taylor_0}
\end{equation}
We assume that $w(s)\simeq 0$ for $s>t$, so that the integration can be extended from 0 to $\infty$ instead of 0 to $t$.

After the integration, in the first term the normalization condition of $w(d)$ can be applied. The second term vanishes, and in the third term the definition of standard deviation $\sigma$ of $w(s)$ can be applied. Eq.~\ref{eq:rt_cont_taylor_0} becomes:
\begin{equation}
    i(t) \simeq \rho(t)\left[ i(t-g) + \frac{\sigma^2}{2}i^{\prime\prime}(t-g) \right]\,.
    \label{eq:rt_cont_taylor}
\end{equation}

If we assume again $i(t) = A\,e^{\lambda t}$, hence $i^{\prime\prime}(t) =A\,\lambda^2\,e^{\lambda t}$, Eq.~\ref{eq:rt_cont_taylor}, becomes:
\begin{equation}
    A\,e^{\lambda t} = \rho(t) \left[A\, e^{\lambda t}e^{-\lambda g} + A\,\frac{\sigma^2}{2}\lambda^2 e^{\lambda t}e^{-\lambda g}\right]\,.
\end{equation}
The term $A\,e^{\lambda t}$ simplifies.
If $\lambda^2\sigma^2 \ll 1$, we may write, approximately:
\begin{equation}
    1 = \rho(t)\,e^{-\lambda g}\left(1+\frac{\sigma^2}{2}\lambda^2\right) \simeq \rho\,e^{-\lambda g} e^{\lambda^2\sigma^2/2}\,, 
\end{equation}
hence:
\begin{equation}
 \rho = e^{\lambda g - \lambda^2\sigma^2/2}\,.
 \label{eq:rho_lambda_lambda2}
\end{equation}

Equation~\ref{eq:rho_lambda_lambda2} has already been reported in~\cite{wallinga_lipsitch}. This result implies the width of the distribution $w_s$ has the effect to replace $g$ in Eq.~\ref{eq:rho_e_lambda_g} with an ``effective'' generation time $g^{\mathrm{eff}}$ that is somewhat smaller than the true average value and depends on $\lambda$ according to:
\begin{equation}
    g^{\mathrm{eff}} = g - \lambda\frac{\sigma^2}{2}\,.
    \label{eq:rt_lista_2}
\end{equation}

In order to take into account more details of the distribution, more terms may be added to Eq.~\ref{eq:rt_cont_taylor_0}. Those would add a dependency of $\rho$ on the higher moments: skewness, kurtosis and possibly more, if required by the desired accuracy. Those cases are not considered in the present work.

\section{``Exact'' solution}
If we assume, as in the previous section, that $i(t)$ is an exponential function, or at least that it can be approximated to an exponential function within a time interval that is at least as wide as the time range in which $w(s)$ is not negligible, $\rho(t)$ can be computed ``exactly'', and is constant within that interval.

If we assume $i(t) = A\, e^{\lambda t}$, Eq.~\ref{eq:it_continuous} becomes:
\begin{equation}
  A\,e^{\lambda t} = \rho(t)\int_0^t A\,e^{\lambda (t-s)}w(s)\,\mathrm{d}s
 = A\,e^{\lambda t} \rho(t)\int_0^t e^{-\lambda s}w(s)\,\mathrm{d}s \,.  \label{eq:it_continuous_exp}
\end{equation}
Simplifying the term $A\, e^{\lambda t}$, as in the previous cases, $\rho(t)$ can be computed as:
\begin{equation}
    \rho(t) = \left[\int_0^t e^{-\lambda s}w(s)\,\mathrm{d}s\right]^{-1}\,.
\end{equation}
If $w(s)$ is negligible for values of $s>t$, we can extend the integration from 0 to $\infty$, and $\rho(t)=\rho$ does not depend on $t$:
\begin{equation}
    \rho = \left[\int_0^{\infty} e^{-\lambda s}w(s)\,\mathrm{d}s\right]^{-1}\,.
\label{eq:rho_exact_const}
\end{equation}
This result is also reported in~\cite{wallinga_lipsitch}.

Note that if $\lambda=0$, Eq.~\ref{eq:rho_exact_const} becomes:
\begin{equation}
    \rho = \left[\int_0^{\infty}w(s) \mathrm{d}s\right]^{-1}\,.
\end{equation}
The normalization of $w(s)$ implies $\rho=1$, regardless of the details of the probability distribution $w(s)$.

\section{The case of a gamma distribution}

In~\cite{cereda}, $w(s)$ is approximated to a gamma distribution:
\begin{equation}
    w(s) = \frac{s^{\kappa-1}e^{-s/\theta}}{\theta^\kappa\Gamma(\kappa)}\,,
\end{equation}
where $\kappa$ and $\theta$, the shape and scale parameters, are determined with a fit to the Italian data.
Equation~\ref{eq:rho_exact_const} becomes:
\begin{equation}
    \rho = {\theta^\kappa\Gamma(\kappa)}\left[\int_0^{\infty} s^{\kappa-1}e^{-s(\lambda +1/\theta)}\,\mathrm{d}s\right]^{-1}\,,
\end{equation}
where the integration can be performed analytically:
\begin{equation}
    \rho = {\theta^\kappa\Gamma(\kappa)}\left[\left.
    -\frac{\Gamma(\kappa, (\lambda +1/\theta)s)}{(\lambda+1/\theta)^\kappa}
    \right|_{s=0}^{s=\infty}\right]^{-1}\,.
\end{equation}
With some simplification of the $\Gamma$ functions, the result is:
\begin{equation}
    \rho = (1+\lambda\theta)^\kappa\,.
    \label{eq:rt_lista_3}
\end{equation}
The above equation is valid for $-1/\theta <\lambda < \infty$.
Again, $\lambda=0$ corresponds to $\rho=1$ for any values of $\kappa$ and $\theta$, as demonstrated in general in the previous section.

\section{$R_t$ and $\tau_2$ as indicators of the epidemic evolution}
$R_t$ is often used as indicator of the epidemic evolution. As we have seen, there is a very close relation between the Effective Reproduction Number and doubling time.
The estimate of the doubling time $\tau_2$ can be determine directly from the number of infected people, while $R_t$ also requires an estimate of the average generation time $g$, which propagates an extra uncertainty with respect to the estimate of $\tau_2$. 

The main feature of $R_t$ is the passage through the threshold value of one: $R_t>1$ indicates a growing phase, while $R_t<1$ indicates a receding phase of the epidemic. Those conditions are equivalent to $\tau_2>0$ and $\tau_2<0$, respectively, as evident form Eq.~\ref{eq:rt_lista}. 
In the case $\hat{\lambda}=0$, $\hat{R}_t$ is not affected by the uncertainty on the estimate of $g$.

For this reason, we consider $\tau_2$, or equivalently $\lambda$, a better indicator of the situation of the epidemic compared to $R_t$, which may be of interest for other epidemiology purposes.

\section{Results}

\begin{figure}[htbp]
    \centering
    \includegraphics[width=0.9\textwidth]{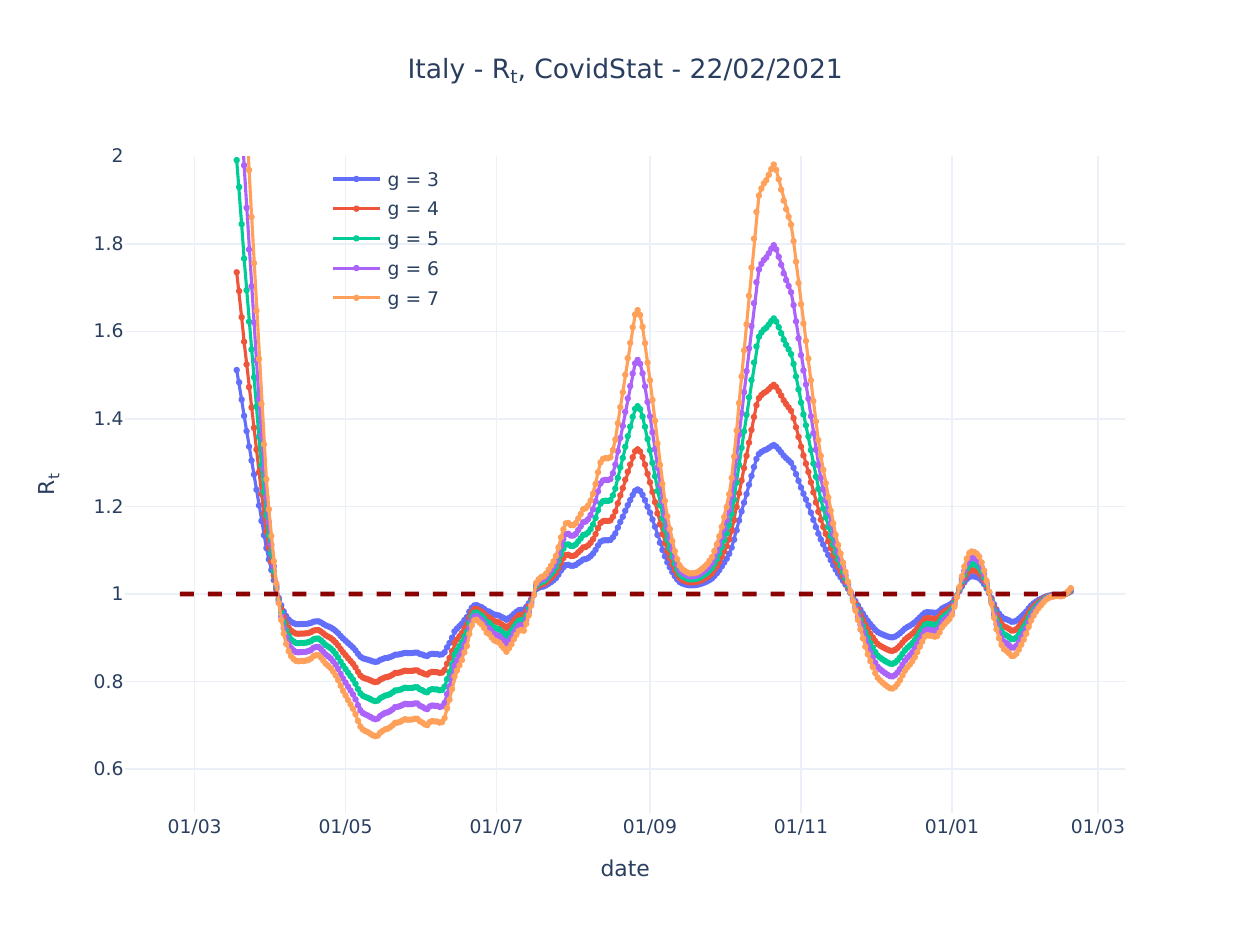}
    \caption{$R_t$ evaluated on the public COVID-19 Italian data released by the Italian Dipartimento di Protezione Civile with the presented algorithm assuming different constant values of the generation time $g$ from 3 to 7 days.}
    \label{fig:rt_covidstat_g}
\end{figure}

Figure~\ref{fig:rt_covidstat_g} shows $R_t$, evaluated with the presented algorithm assuming a constant generation time, using the public Italian COVID-19 data released by the Italian Dipartimento di Protezione Civile~\cite{dpc}. Different values of the average generation time $g$ have been assumed, from 3 to 7 days.

 The magnitude of the dependence of $R_t$ on $g$ gives also a clue about the uncertainty on $R_t$ due to imperfect knowledge of $g$, which mainly affects the regions where $R_t$ is significantly different from 1.
 
\begin{figure}[htbp]
    \centering
    \includegraphics[width=0.9\textwidth]{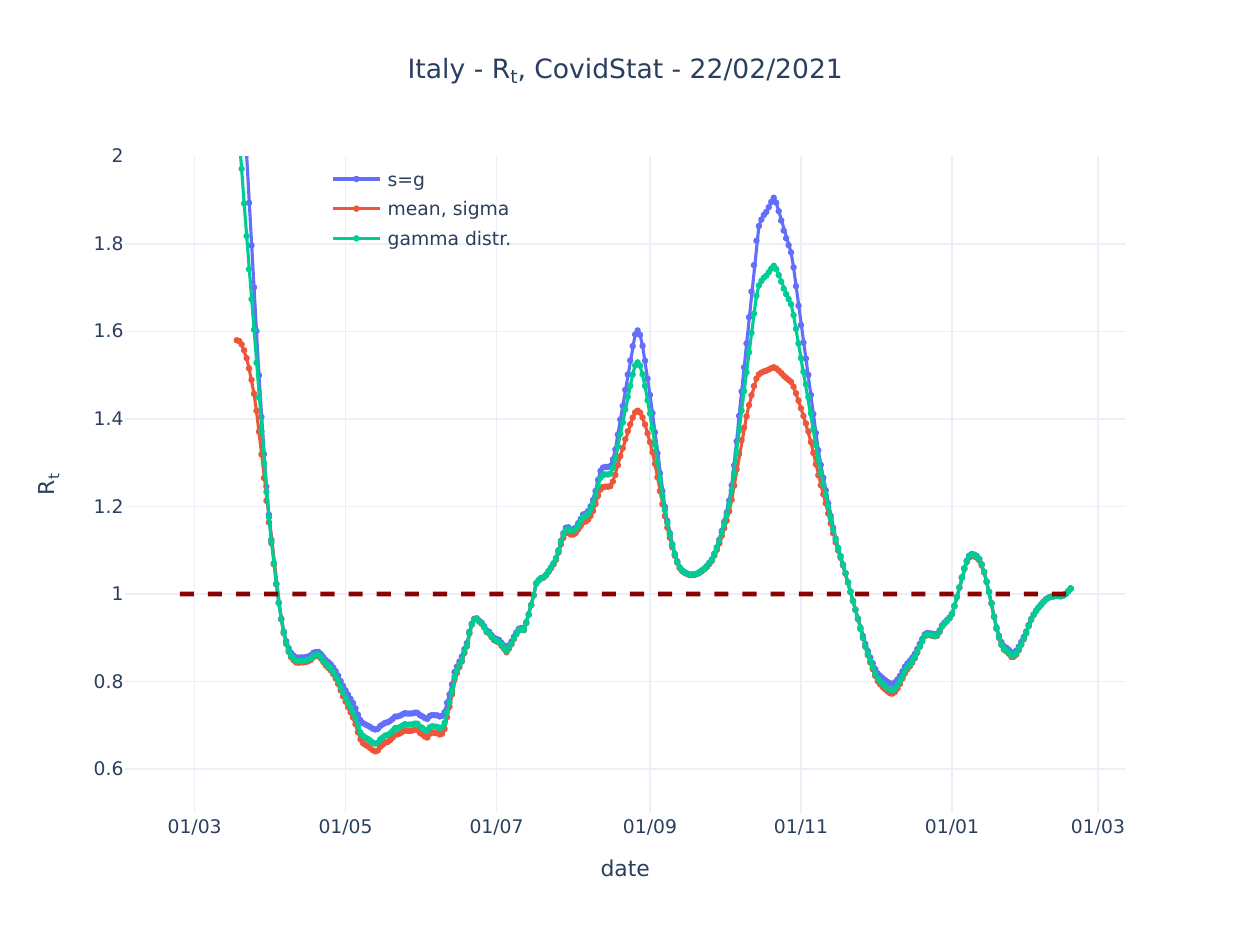}
    \caption{$R_t$ evaluated on the public COVID-19 Italian data released by the Italian Dipartimento di Protezione Civile with assuming a constant generation time, assuming a mean value and a standard deviation contribution, and assuming a gamma distribution. The assumed parameters are taken from~\cite{cereda}.
    }
    \label{fig:rt_covidstat_comp}
\end{figure}

 Figure~\ref{fig:rt_covidstat_comp} shows instead the evaluation performed with the three models discussed above:
 \begin{enumerate}
     \item Eq.~\ref{eq:rt_lista}, assuming a constant generation time of $g=6.7$ days;
     \item Eq.~\ref{eq:rt_lista_2}, assuming a mean value of 6.7 days and a standard deviation of 4.88 days;
     \item Eq.~\ref{eq:rt_lista_3}, assuming a gamma distribution having parameters $\kappa=1.87$ and $\theta=3.57$ days, respectively, as determined in~\cite{cereda}
  \end{enumerate}
 Note that the mean of the gamma distribution is equal to the product $\kappa\theta$.
 
All three methods give similar values for $R_t$ close to 1, but exhibit some discrepancy at more extreme values. Compared to the ``exact'' solution that assumes a gamma distribution (Eq.~\ref{eq:rt_lista_3}), assuming a fixed generation time (Eq.~\ref{eq:rt_lista}) gives a result that is about 9\% larger at the highest value and about 4\% larger at the lowest value. Including the contribution of the standard deviation term (Eq.~\ref{eq:rt_lista_2}) gives a reduction of about 12\% at the larges value and 3\%  at the lowest value. Using (Eq.~\ref{eq:rt_lista}) with a lower ``effective'' $g$ may improve the agreement with the ``exact'' solution at higher values at the cost of a poorer agreement at lower values. This is effectively done in the implementation of the RKI algorithm.

\begin{figure}[htbp]
    \centering
    \includegraphics[width=0.9\textwidth]{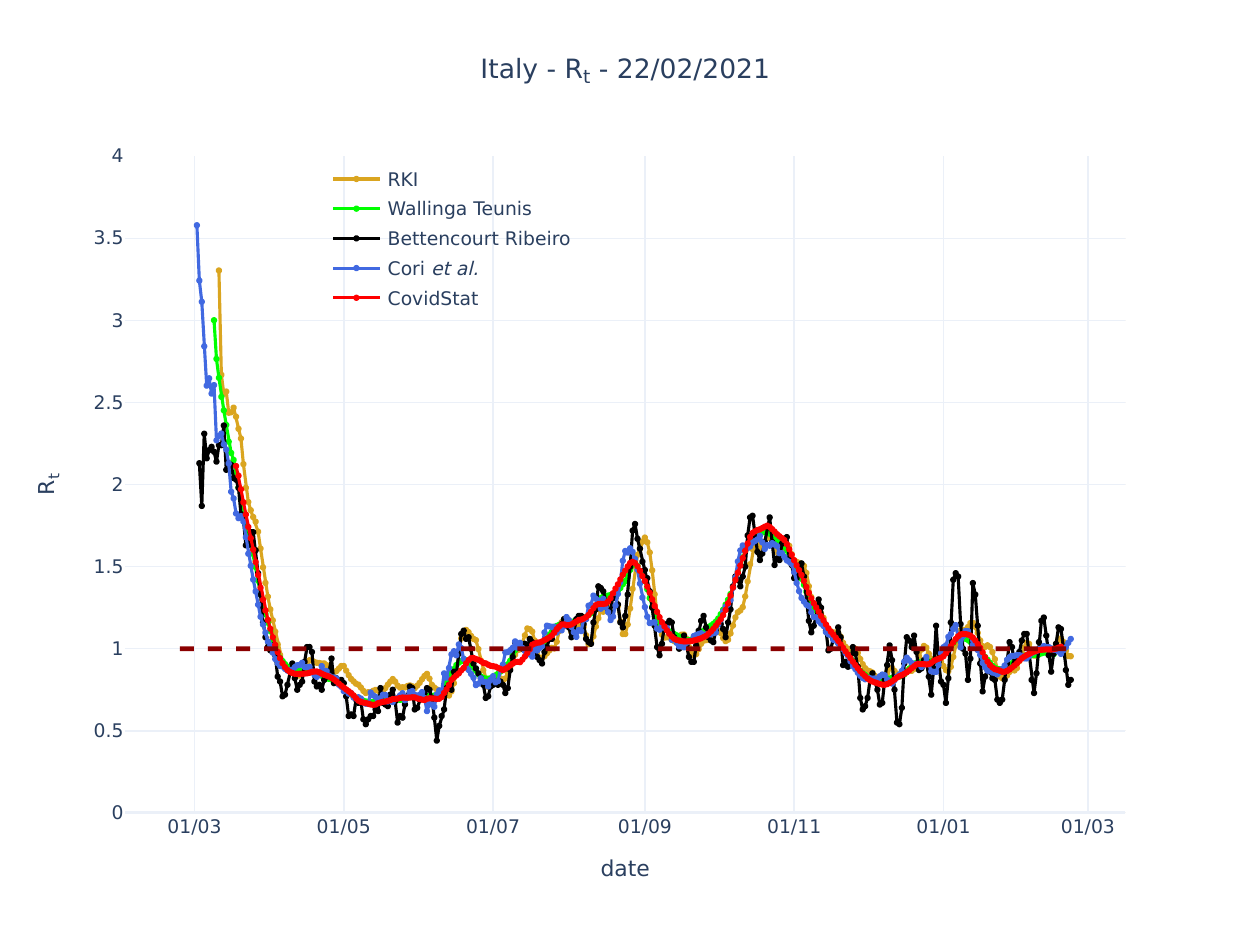}
    \caption {Comparison of $R_t$ computed using different algorithms with public COVID-19 Italian data from the Italian Dipartimento di Protezione Civile. The algorithm presented in this paper is noted as CovidStat and assumes a gamma distribution with known parameters. It is compared with algorithms by RKI, Wallinga and Teunis, Bettencourt and Ribeiro, and Cori {\it et al.}.}
    \label{fig:rt_italy}
\end{figure}

Figure~\ref{fig:rt_italy} shows the application of different algorithms to the official Italian COVID-19 data published by the Italian Dipartimento di Protezione Civile~\cite{dpc}. The algorithm presented in this paper is noted as CovidStat and assumes a gamma distribution with the parameters reported above. It is compared with algorithms by Wallinga and Teunis~\cite{wallinga},  Bettencourt and Ribeiro~\cite{bettencourt}, Cori {\it et al.}~\cite{cori}, and RKI~\cite{rki}.
Algorithms by  Wallinga and Teunis and Cori {\it et al.} use the details of the probability distribution $w_s$ and are here implemented assuming the same $w_s$ as our algorithm. Bettencourt and Ribeiro uses a fixed time, that we have set to 7 days.

The method proposed here has been implemented with an exponential fit to the last 14 days. The RKI algorithm has been applied with generation time $g=5$, since the original implementation with $g=4$ showed significant discrepancy with respect to the other algorithms, consistently with what can be noted in Fig~\ref{fig:rt_covidstat_comp}. A smoothing of the infection data with a Savitzky-Golay filter~\cite{savitzky} using a time window of 15 days and a third-order polynomial was also applied to the infection data before applying the RKI algorithm.

\begin{figure}[htbp]
    \centering
    \includegraphics[width=0.9\textwidth]{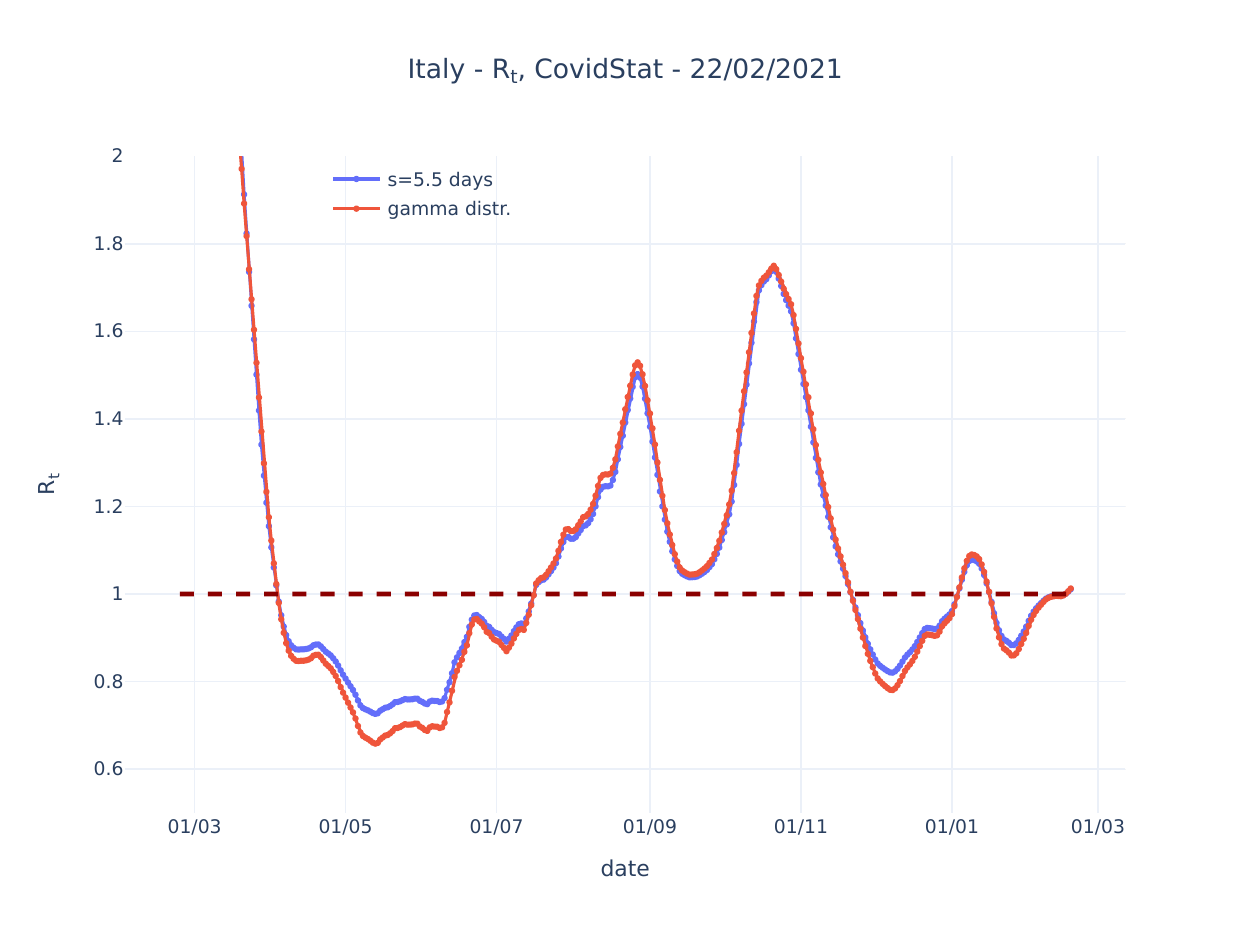}
    \caption {Comparison of $R_t$ computed assuming a gamma distribution and assuming a constant generation time ``tuned'' to $g=5.5$ in order to reduce the disagreement for $R_t>1$. A residual disagreement for $R_t<1$ is visible.}
    \label{fig:rt_comp_2}
\end{figure}

The comparison of the proposed method with other algorithms shows a good agreement, considering the possible source of uncertainties and the intrinsic ``ripple'' structure of the data that may depend on the applied smoothing. In particular, agreement of our method is very good with the Wallinga-Teunis and with the Cori {\it et al.} algorithms. The agreement with the Bettencourt-Ribeiro is also good, considering that it includes a ``ripple'' structure due to the data fluctuations. The agreement with the RKI method is also reasonable after the assumed constant generation time is ``tuned'', with a residual disagreement for the cases where $R_t<1$. This feature is consistent with what can been observed comparing the ``exact'' solution computed for the gamma distribution to the one computed assuming a fixed generation time ``tuned'' to the more convenient value $g=5.5$, as shown in Fig~\ref{fig:rt_comp_2}.

\begin{figure}[htbp]
    \centering
    \includegraphics[width=0.9\textwidth]{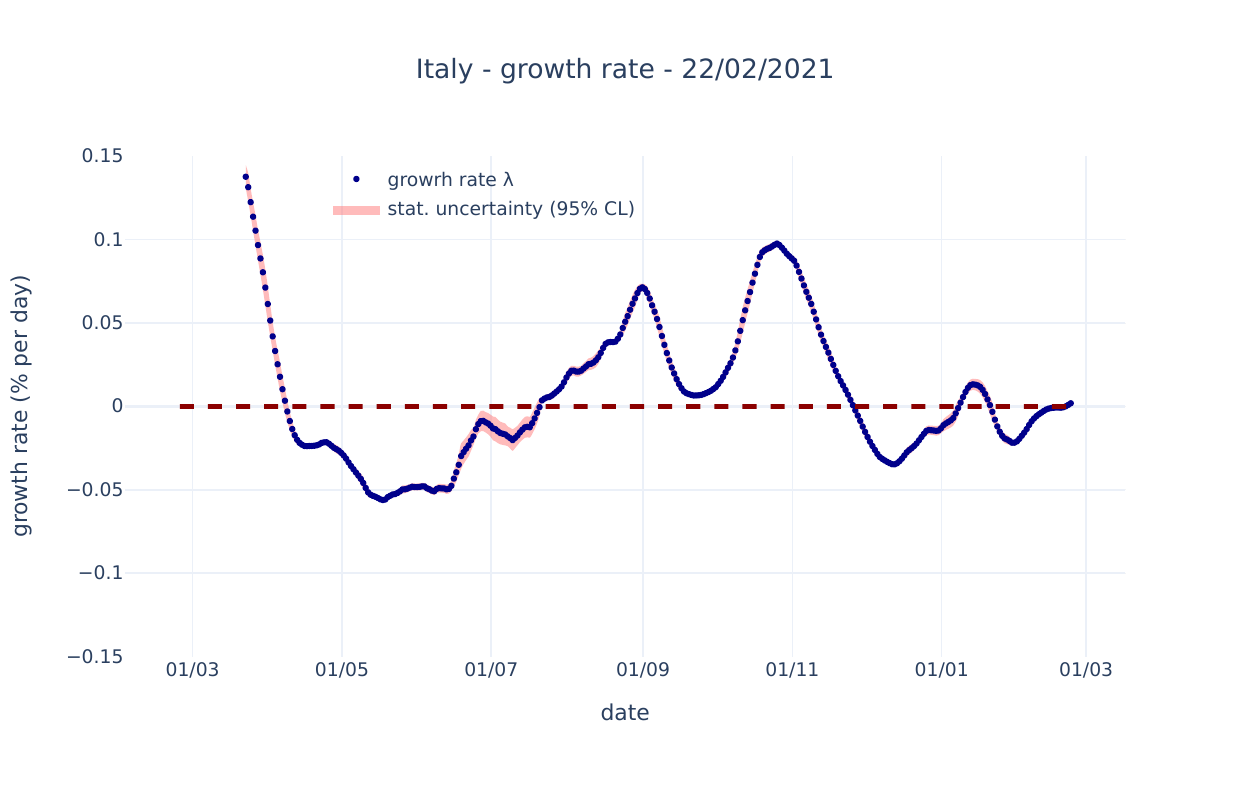}
    \includegraphics[width=0.9\textwidth]{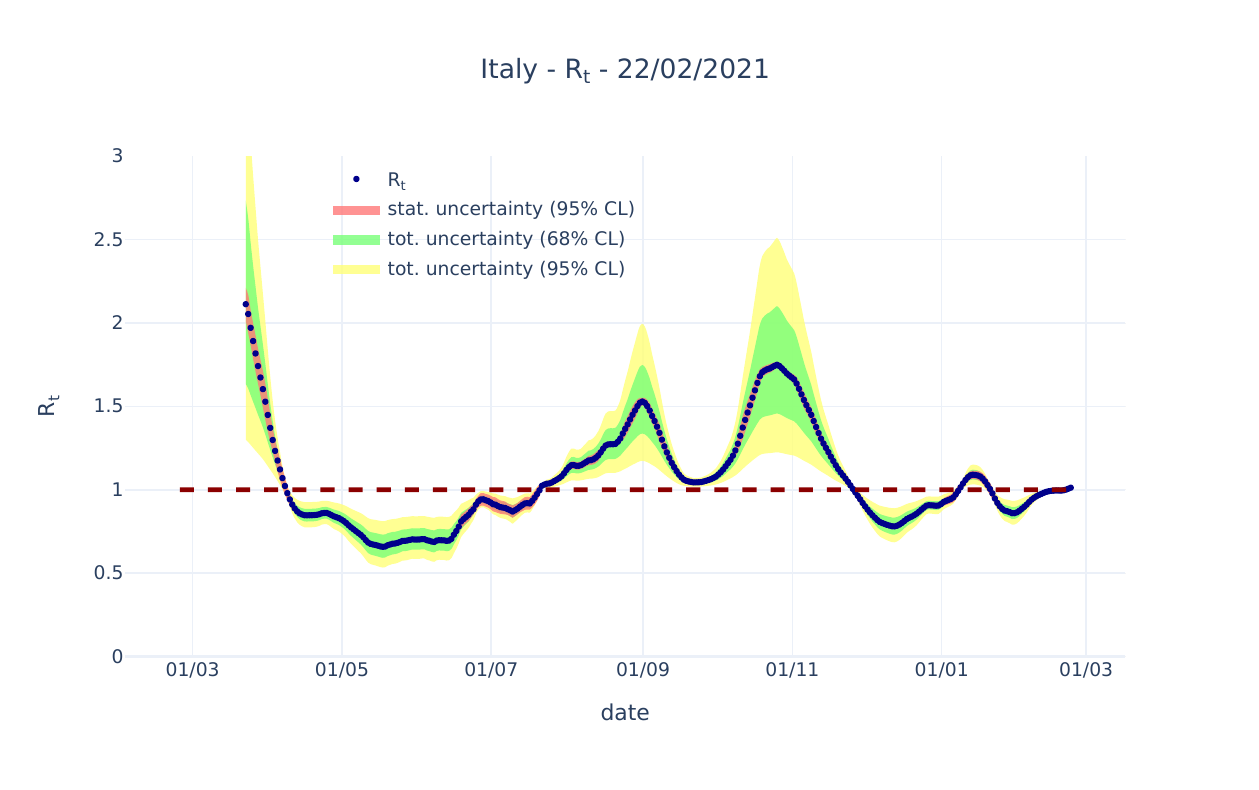}
    \caption {Growth rate $\lambda$ (top) and $R_t$ (bottom).
     For the growth rate $\lambda$, the statistical uncertainty band at 95\% Confidence Level is shown.
     For $R_t$ the contribution to uncertainty due to the propagation
     of the statistical uncertainty on $\lambda$ at 95\% Confidence Level is shown together with the total uncertainty at the 68\% and 95\% 
     Confidence Level, that also takes in to account the uncertainty on the parameters that model $w(s)$.
     All data refer to Italy according to public COVID-19 Italian data from the Dipartimento di Protezione Civile.
     }
    \label{fig:italy_gr}
\end{figure}

Figure~\ref{fig:italy_gr} shows the estimated growth rate $\lambda$ and the corresponding $R_t$ for Italy data. Estimates are done with an exponential fit over the last 14 days. For $R_t$ the contribution to uncertainty due to the propagation of the statistical uncertainty on $\lambda$ is, in most of the range, much smaller than the total uncertainty that also takes in to account the uncertainty on the parameters that model $w(s)$, according to the estimate from~\cite{cereda}. This contribution to the total uncertainty is particularly large as the values of $R_t$ depart from one. For $R_t=1$, as noted before, the uncertainty contribution form the parameters that model $w(s)$ is null. The magnitude of the total uncertainty is comparable with what is obtained from the algorithm by Cori {\it et al.} that tales into account the uncertainty on $w(s)$.

\section{Performances}

We compared the CPU time required to run the five algorithms considered in this paper. The benchmarks ran on a dedicated cluster with 32 cores/64 threads on two AMD EPYC 7301 processors and 64GB RAM. The algorithm ran on a single thread avoiding any multithread implementation. The results are reported in Table~\ref{tab:cpu}.

\begin{table}[htbp]
\begin{center}
\begin{tabular}{| l | r | r | r | r | r | r |}
\hline
 \multicolumn{1}{|c}{Geographic}  &  \multicolumn{1}{|c}{Inhabitants} &  \multicolumn{1}{|c}{\multirow{2}{*}{CovidStat}} &  \multicolumn{1}{|c}{Wallinga}  &  \multicolumn{1}{|c}{Bettencourt}  
 &  \multicolumn{1}{|c}{Cori} &  \multicolumn{1}{|c|}{\multirow{2}{*}{RKI}}   \\ 
    \multicolumn{1}{|c}{area} &  \multicolumn{1}{|c}{(mln., 2019)} & \multicolumn{1}{|c}{} &   \multicolumn{1}{|c}{Teunis} &   \multicolumn{1}{|c}{Ribeiro} 
 &  \multicolumn{1}{|c|}{\it et al.} &   \\ \hline
 Italy & 60.36 &  0.138 & 510.1 & 80.3 & 11.5 & 0.587 \\  
 Lombardia & 10.06 & 0.120 & 80.8 & 78.3 & 11.5 & 0.585 \\ 
 Lazio & 5.88 & 0.105 & 30.0 & 76.1 & 11.2 & 0.586 \\  
 Campania & 5.80 & 0.103 & 31.9 & 33.9 & 11.2 & 0.584 \\  
 Emilia-Romagna & 4.46 & 0.101 & 33.5 & 75.2 & 11.4 & 0.585 \\  
 Basilicata & 0.56 & 0.098 & 20.1 & 28.9 & 11.3 & 0.584 \\  
\hline
\end{tabular}
\caption {\label{tab:cpu}
CPU time in seconds required to run the five $R_t$ algorithms for Italy and five Italian regions with decreasing number of inhabitants and the number of infected persons. Emilia-Romagna has lower number of inhabitants, but significantly more infected persons compared to Lazio and Campania.
The specs of the cluster used for the benchmark are reported in the text.}
\end{center}
\end{table}

 The algorithm proposed in this paper outperforms all other algorithms, in particular when the number of cases is large, as for Itay and Lombardia. The comparison with the RKI algorithm is not very meaningful. RKI estimates $R_t$ as the ratio of the number of infected persons last $g=5$ days divided by the the number of infected persons in the previous $g$ days, which takes a very small CPU time.  
 Nonetheless, our implementation is largely dominated by the overhead introduced by the python module {\tt pandas}~\cite{pandas} compared to {\tt numpy}~\cite{numpy}, which is faster, and is the one we use for the CovidStat algorithm. The choice was only dictated by convenience, and we didn't consider any porting of our implementation of the RKI algorithm to {\tt numpy}, that would outperform the CovidStat algorithm, because the gain would be negligible anyway.

We report in the CovidStat website~\cite{our web site} $R_t$ estimates for Italy, for North, Center and South separately, for the 20 Italian regions, and for the autonomous provinces of Bolzano and Trento. On the aforementioned dedicated 64-thread cluster, each geographic area running on a separate thread, the computation takes about 30 minutes for all five algorithms, including all the data management overhead. 

In addition, we compute $R_t$ for the 107 provinces and for about 30 countries. For those, we only compute the CovidStat $R_t$ estimate in order to reduce the required computation time. This evaluation takes a negligible CPU compared with the other methods of computation of $R_t$ computations.

Updates are published on our website daily and are produced automatically, with no human intervention, as soon as the data from the Dipartimento della Protezione Civile are available.

\section{Conclusion}
A simplified method to determine an estimate of $R_t$ based on a local exponential fit is presented. The method can be applied assuming a fixed generation time, including the contribution of the standard deviation of the generation time distribution, or assuming a functional form for the probability distribution of the generation time. If a gamma distribution is assumed, a simple analytic solution is reported.
The method offers some advantages compared to the simplified method adopted by the Robert Koch Institut, yet preserving good computing performances that makes it suitable for a real-time evaluation. 

Results of the method applied to the public Italian COVID-19 data have been presented. The proposed method shows a good agreement with other, more complex, algorithms available in literature and implemented in public software packages.

We note a close relation between $R_t$ and the doubling time of the number of infections $\tau_2$, or equivalently the growth rate $\lambda$. In particular, the condition $R_t>1$ is equivalent to $\tau_2>0$ or $\lambda>0$. Since the determination of $R_t$ is affected by additional uncertainty sources compared to $\tau_2$, we consider $\tau_2$ or $\lambda$ to be a more sound and simpler indicator of the condition of growing or receding epidemic compared to $R_t$, while $R_t$ may have more importance in other contexts of epidemiological interest.

We publish in real time daily estimates of $R_t$ as computed by our algorithm and by all the other algorithms quoted in this article for the cases in Italy and all the Italian regions under \cite{our web site}. Daily values for the major world countries are also reported.

\section{Acknowledgement}
The present work has been done in the context of the INFN CovidStat project that produces an analysis of the public Italian COVID-19 data. The results of the analysis are published and updated daily on the website 
{\tt covid19.infn.it/}. The project has been supported in various ways by a number of people from different INFN Units. In particular, we wish to thank, in alphabetic order: Stefano Antonelli (CNAF), Fabio Bredo (Padova Unit), Luca Carbone (Milano-Bicocca Unit), Francesca Cuicchio (Communication Office), Mauro Dinardo (Milano-Bicocca Unit), Paolo Dini (Milano-Bicocca Unit), Rosario Esposito (Naples Unit), Stefano Longo (CNAF), and Stefano Zani (CNAF). We also wish to thank Prof. Domenico Ursino (Universit\`a Politecnica delle Marche) for his supportive contribution.

\end{document}